\documentclass[aps,prl,showpacs,twocolumn,showkeys,amssymb,amsmath,nobibnotes,nofootinbib,superscriptaddress,
reprint]{revtex4}
\usepackage{verbatim}
\usepackage{amssymb}
\usepackage{graphicx}
\usepackage{subfigure}
\usepackage{epstopdf}
\usepackage[colorlinks=true,linkcolor=red]{hyperref}
\usepackage[sort&compress]{natbib}
\usepackage{color}
\usepackage{changes}
\usepackage[toc,page]{appendix}

\begin{document}

\title{Kick-induced rectified current in symmetric nano-electromechanical shuttle}

\author{Pinquan Qin} 
\affiliation{Department of Physics, Wuhan University of Technology, Wuhan, 430070, China}
\author{Hee Chul Park}
\email{hcpark@ibs.re.kr}
\affiliation{Center for Theoretical Physics of Complex Systems, Institute for Basic Science (IBS), Daejeon, Korea, 34126}

\begin{abstract}

We have studied the rectified current in a geometrically symmetric nano-electromechanical shuttle with periodic kicks and sinusoidal ac bias voltages. 
The rectified current is exactly zero under the geometrical symmetry which is generated by the electrons transferred from source to drain electrodes through the movable shuttle.
We investigate the nonzero rectified currents through the symmetric shuttle with regular motion of which the time-translational symmetry is broken. The motion of the shuttle, moreover, becomes chaotic with the same mechanism of the kicked rotor and generates the scattered current as increasing kick strength. 
We point out that the time-translational-symmetry breaking of the instantaneous current is an important role of manipulation of the rectified current.

\end{abstract}

\maketitle

\subsection*{Introduction}
The rapid development of nanotechnology in recent years has created a new quantum electronic devices which incorporate the mechanical degree of freedom. 
This device is called nano-electromechanical systems (NEMS)\cite{cleland2013foundations,lyshevski2018mems}. 
In this system, a movable mesoscopic object, namely, the single-electron shuttle, begins to transport electrons one by one beyond a certain critical bias \cite{gorelik1998shuttle, chulki17collide, rong2013transition,moeckel2014synchronizing,kulinich2014single, zhao2016stochastic, kim2013realizing,kim2015tracing}. 
Here, the Coulomb energy and mechanical motion make Ohm's law invalid for the electronic conductance so that the current is not proportional to the voltage drop across the devices. Instead of the Coulomb blockade, the current is discretized proportional to the number of electrons and vibration frequency. 
Since quantum mechanical tunneling probability is exponentially
sensitive to the tunneling distance between the shuttle and electrodes, the position of the shuttle is crucial to define the system.
On the other hand, the Coulomb force which accompanies the discrete nanoscale charge fluctuations drives the motion of the shuttle. 
Previous research \cite{pistolesi2005charge} reported a geometrically symmetric shuttle with sinusoidal ac driven voltage has an exactly zero rectified current. We want to treat the case of the symmetric shuttle with the periodic driven forces beyond simply sinusoidal ac voltage.

The kicked rotor model is a prototypical model to study classical or quantum periodic systems including nonlinearity and chaos \cite{chirikov1979universal, casati1979stochastic,izrailev1990simple}.
The kicked rotor model is a single particle system with a kicked driven potential of which the strengths are discrete at every periodic time $T_k$.
In the classical theory, the particle with a strong kick strength shows the chaotic motion with a positive Lyapunov exponent.
In the quantum theory, two momentum space extraordinary phenomena arise: dynamic localization \cite{casati1979stochastic,bayfield1989localization,moore1994observation,bitter2016experimental} and quantum resonance \cite{izrailev1980quantum,sadgrove2004observation,dana2008experimental}, which correspond to whether the driving period is an irrational or rational multiple of $2\pi$, respectively.  
There are rectified currents in the geometrically asymmetric nano-electromechanical shuttle with time-periodic bias voltage \citep{pistolesi2005charge, pqqin2019}.
The results enlarge the knowledge about the dynamics of self-excited oscillator in nanoscale, and provide available ways to optimize the nano-devices to generate rectified current with practical purpose.    

\begin{figure}
  \centering
  \includegraphics[width=.9\linewidth]{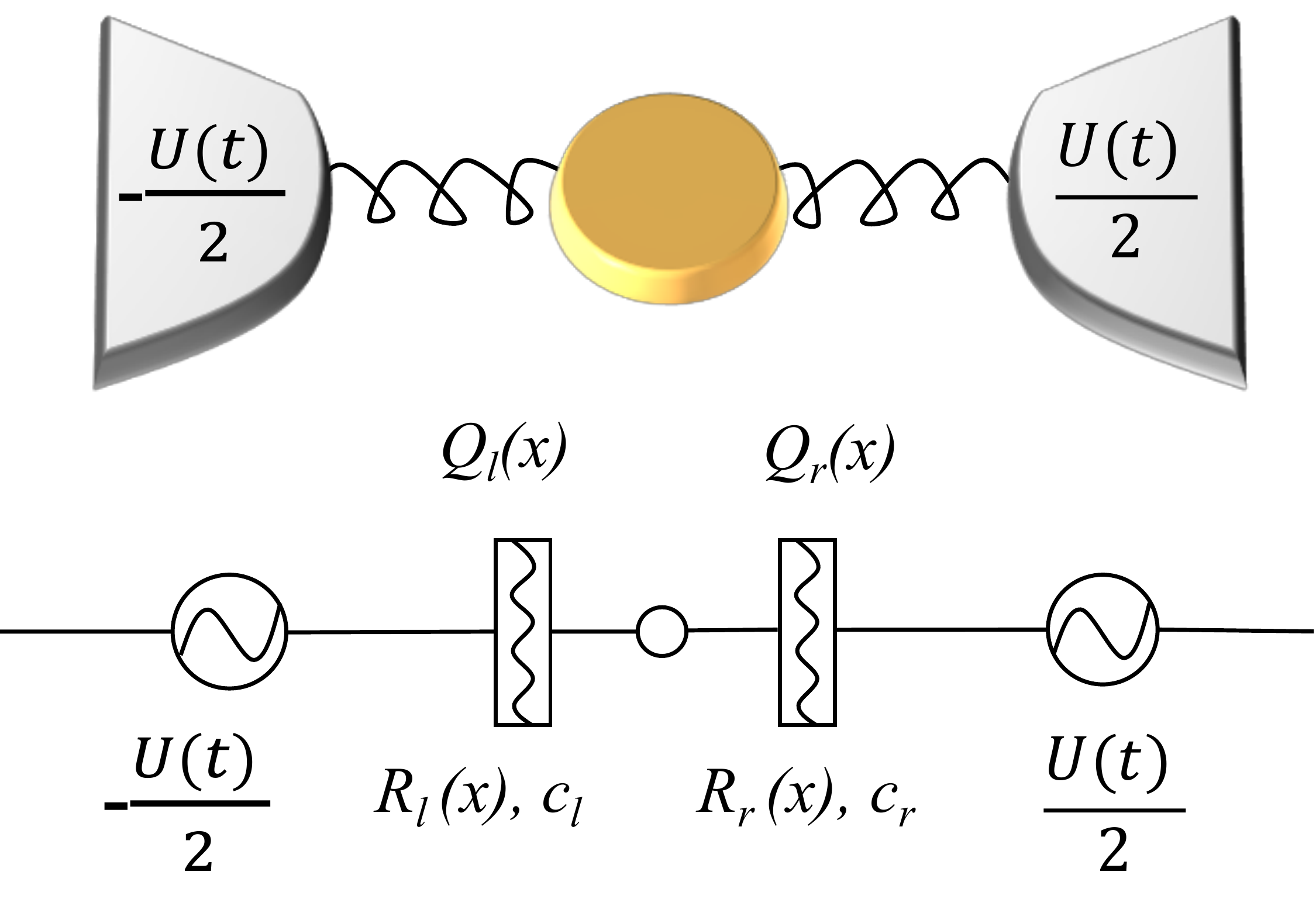}
  \caption{\label{structure}
A cartoon shows an electron shuttle with symmetric bias voltages and the equivalent circuit is drown below. The combination of rectangular box and wavy line in the circuit represents position-dependent tunnel junction. }
\end{figure}

In this research, we study a geometrically symmetric electron shuttle with periodic bias voltage which is the combination of kicks and sinusoidal ac voltage. 
We find that the displacement of the symmetric shuttle draws the phase diagram so called Arnold's tongues indicating instabilities and that the rectified current is nonzero in regular motion when we turn on the kicks.
In addition to the regular motions, we also observe a chaotic motion of the shuttle with increasing kicks strength.
This phenomenon is caused by the nonlinear force induced by the  bias voltage as kicks, even though the system is exactly geometrically symmetric. 
It is noticed that the rectified current is generated by the time-translational-symmetry breaking of the instantaneous current due to the interplay between self-oscillation period and external kick period. 
Moreover, the scattered rectified currents are induced by the chaotic motion of the shuttle due to the mechanism of kicked harmonic oscillator.


\subsection*{Model of a symmetric shuttle}
Let's consider a nano-electromechanical shuttle which is the combination of two metallic electrodes and a movable nano-dot. 
The movable nano-dot is initially and symmetrically located in the center of two electrodes. 
The distance between two electrodes is large enough to make the nano-dot (or nano-shuttle) oscillating sufficiently without touching any electrode.
We apply a symmetric and time-periodic voltage to the electrodes on both sides (see Fig.\ref{structure}). 
The applied voltage drives the shuttle oscillating between two electrodes and the force is a function of $x$ which indicates the displacement of the shuttle. We assume that the pumped current and current-induced force contributed by nonequilibrium are negligible \cite{reckermann2010interaction,calvo2012interaction,bustos2019thermodynamics}. 

In the adiabatic limit, namely, the electronic relaxation is much faster than the mechanical motion, the classical circuit shown in Fig \ref{structure} can be used to analyze electronic properties of the shuttle \cite{ahn2006current}.
Base on Kirchhoff's law of a circuit structure, we obtain the following equations:  
\begin{eqnarray}
\frac{Q_l(x)}{R_l(x)c_l} - \frac{Q_r(x)}{R_r(x)c_r} &=& 0,\label{qrc}\\
\frac{Q_l(x)}{c_l} + \frac{Q_r(x)}{c_r} &=& U(t)\label{qvx}
\end{eqnarray}
where $Q_j$ is the charge, $R_j$ is the resistance, $c_j$ is the capacitance at the $j$-th junction, and $U$ is the applied time-dependent voltage. The position-dependent resistances at the each junction are $R_{l}=R_l^0e^{(d+x)/\lambda}$ and $R_{r} = R_{r}^0 e^{(d-x)/\lambda}$ with the consideration of the electron tunneling process between the electrodes and the shuttle, $d$ is the half distance between two electrodes, and $\lambda$ is the phenomenological tunneling length.
Under the geometrical symmetry, $R_l^0e^{d/\lambda}=R_r^0e^{d/\lambda}=R_0$ and $c_l=c_r=c$, position-dependent resistances of left and right electrodes are $R_{l}=R_0 e^{x/\lambda}$ and $R_{r}=R_0 e^{-x/\lambda}$.
The total charge amount on the movable nano-shuttle as follow,
\begin{eqnarray}
Q_s(x,t)=&cU(t)\frac{R_{l}(x)-R_{r}(x)}{R_l(x)+R_r(x)} \nonumber \\
=&cU(t)\tanh(x/\lambda).
\label{qt}
\end{eqnarray}

The time-dependent driven bias voltage between two leads is the combination of the normal sinusoidal ac voltage and time-periodic kicks which is the pulsed driven voltage, $U(t)=\tilde{U}(t)+\xi D_{\varepsilon}(t/T_k)$, where $\tilde{U}(t)$ is the normal sinusoidal ac voltage, $\tilde{U}(t)=\alpha\sin\omega t$.
$\alpha$ and $\xi$ are the strength of sinusoidal ac voltage and kicks, respectively. 
Discrete kicks only affect on the shuttle during $\varepsilon$ as following,
\begin{eqnarray}
D_{\varepsilon}(t/T_k)=\begin{cases}
0 & t/T_k\in [ n-1,n-\varepsilon),\\
1 & t/T_k\in [ n-\varepsilon,n),
\end{cases}
\end{eqnarray}
where $T_k$ is the kick period, $\varepsilon$ is the kick duration under keeping a very small number $\xi\varepsilon\ll 1$.
The electric energy of the oscillating shuttle is $E(x,t)=Q_s(x,t)U_s(x,t)=Q_s^2(x,t)/2c$, where the voltage drop between two electrodes is $U_{s}(x,t) = Q_s(x,t)/2c$ based on Eq. (\ref{qrc}), (\ref{qvx}) and the symmetric voltage configuration on the electrodes.
Correspondingly, the electric force exerted on the shuttle is as follow, 
\begin{equation} 
F=-\frac{\partial E(x,t)}{\partial x} = -\frac{cU^{2}(t)}{\lambda} \mathcal{F}(x),  
\end{equation} 
where $\mathcal{F}(x)=\sinh (x/\lambda)/\cosh^{3}(x/\lambda)$.
We write $F=F_c+F_k$ which is combined with the smooth and continuous force $F_c = -c\alpha^2\sin^2\omega t\mathcal{F}(x)/\lambda$ induced by the sinusoidal ac voltage and the discrete kicked force $F_k = -c\xi^{2}D_{\varepsilon}(t/T_k)\mathcal{F}(x)/\lambda$ induced by the kicks. 
In order to investigate the influence of $F_k$, we keep a weak coupling and general kick strength as $\alpha\ll\xi$ and $\xi^2\varepsilon$ is finite.

The equation of motion for the nano-shuttle is governed by Newton's equation as a damped harmonic oscillator,
\begin{equation}
m\ddot{x} +m\gamma\dot{x} + m\omega^{2}_{0}x= F_c+F_k,
\label{mequ}
\end{equation}
where $\gamma$ is the dissipation coefficient. The governing equation respects the parity symmetry under $(x\rightarrow -x)$. 
The symmetry is satisfied by both the continuous force and the discrete kicked force.

The instantaneous current between two electrodes is defined by $I(t)=Q_l(x,t)/cR_l(x)=U(t)/2R_0\cosh x/\lambda$ through charge distribution on the shuttle.
Given time interval $[0,\tau]$, the average current is calculated as 
\begin{equation}
I_{dc}=\frac{1}{2R_0 \tau}\int_{0}^{\tau}\frac{\tilde{U}(t)+\xi D_{\varepsilon}(t/T_k)}{\cosh x_{t}/\lambda} dt,
\end{equation}
which is the rectified current with the period $\tau=T_I$.
Under the condition $\xi\varepsilon\ll 1$, we find $\int_{0}^{T_I}\xi D_{\varepsilon}(t/T_k)/\cosh (x_{t}/\lambda) dt \sim 0$. Then the rectified current is reduced as follow,
\begin{equation}
I_{dc}=\frac{\alpha}{2R_0 T_I}\int_{0}^{T_I}\frac{\sin\omega t}{\cosh x_{t}/\lambda} dt.
\label{dcc}
\end{equation}

For the simplification, we use dimensionless equation of motion as
$\ddot{\tilde{x}}+\tilde{\gamma}\dot{\tilde{x}}+\tilde{x}= \tilde{F}_c + \tilde{F}_k$,
where $\tilde{F}_c =  -\tilde{\alpha}^2\sin^2\tilde{\omega}\tilde{t}\mathcal{F}(\tilde{x})$, $\tilde{F}_k =-\tilde{\xi}^{2}\tilde{D}_{\varepsilon}(\tilde{t})\mathcal{F}(\tilde{x})$,
$\mathcal{F}(\tilde{x})=\sinh \tilde{x}/\cosh^{3}(\tilde{x})$.
Here $\tilde{x}=x/\lambda$, $\tilde{t}=t\omega_0$,
$\tilde{\gamma}=\gamma/\omega_0$,
$\tilde{\alpha}=\alpha\sqrt{c}/\sqrt{m}\lambda\omega_0$,
$\tilde{\xi}=\xi\sqrt{c}/\sqrt{m}\lambda\omega_0$,
and $\tilde{\omega}=\omega/\omega_0$.
$\tilde{D}_{\varepsilon}(\tilde{t})\equiv D_{\varepsilon}(\tilde{t}/\tilde{T}_k)$ with period $\tilde{T}_k=T_k\omega_0$.
The corresponding rectified current is calculated as
$I_{dc}=(I_0\tilde{\alpha}/\tilde{\tau})\int_0^{\tilde{\tau}/\omega_0}\sin\tilde{\omega}\tilde{t}/\cosh \tilde{x}d\tilde{t}$,
where $I_0=\lambda\omega_0\sqrt{m}/2\sqrt{c}R_0$ is the magnitude of the current.
In the following discussion, we omit all tilde ( $\tilde{}$ ) symbols.

\subsection*{Driven nano-shuttle without kicks}
\label{sec:orgec5aaac}

We introduce a driven nano-shuttle without kicks to comprehend a simplest case for zero rectified current at the fixed point and limit cycles.
The system is invariant under parity transformation with only sinusoidal ac voltage $\ddot{x} +\gamma\dot{x} + x = F_c$ since the time-dependent part of $F_c$ has the period of $\pi/\omega$ which gives the two equivalent solutions $\pm x$ under geometrical symmetry.
The period of the shuttle is $\pi/\omega$ or $2\pi/\omega$ by considering the Floquet theory and the equation of  motion as a second order differential equation.  
In case of odd period of the shuttle, $\pi/\omega$, any position of the shuttle $x(t)$ is back to the same position after time shift $t\rightarrow t+\pi/\omega$. 
On the other hand, even period of the shuttle, $2\pi/\omega$ has two possibilities $x(t+\pi/\omega)=\pm x(t)$.
For both of cases, the rectified currents show exactly zero because the position dependence of the rectified current is an even function and the time dependence of it is an odd function, $\sin\omega(t+\pi/\omega)=-\sin\omega t$ based on Eq.(\ref{dcc}).
In this situation, the instantaneous current has time-translational symmetry as $I_{t^{\prime}}=-I_t$, where two time points are $t$ and $t^{\prime}=t+\pi/\omega$.
Therefore the rectified current in the shuttle with the geometrical symmetry and sinusoidal ac driven voltage is always zero for periodic regular motion with any $\alpha$ and $\omega$.
This is the common view about the rectified current of the electron shuttle with geometrical symmetry.
In the following content we point out that breaking the time-translational symmetry can generate the finite rectified current.

\begin{figure}
\centering
\includegraphics[width=1.0\linewidth]{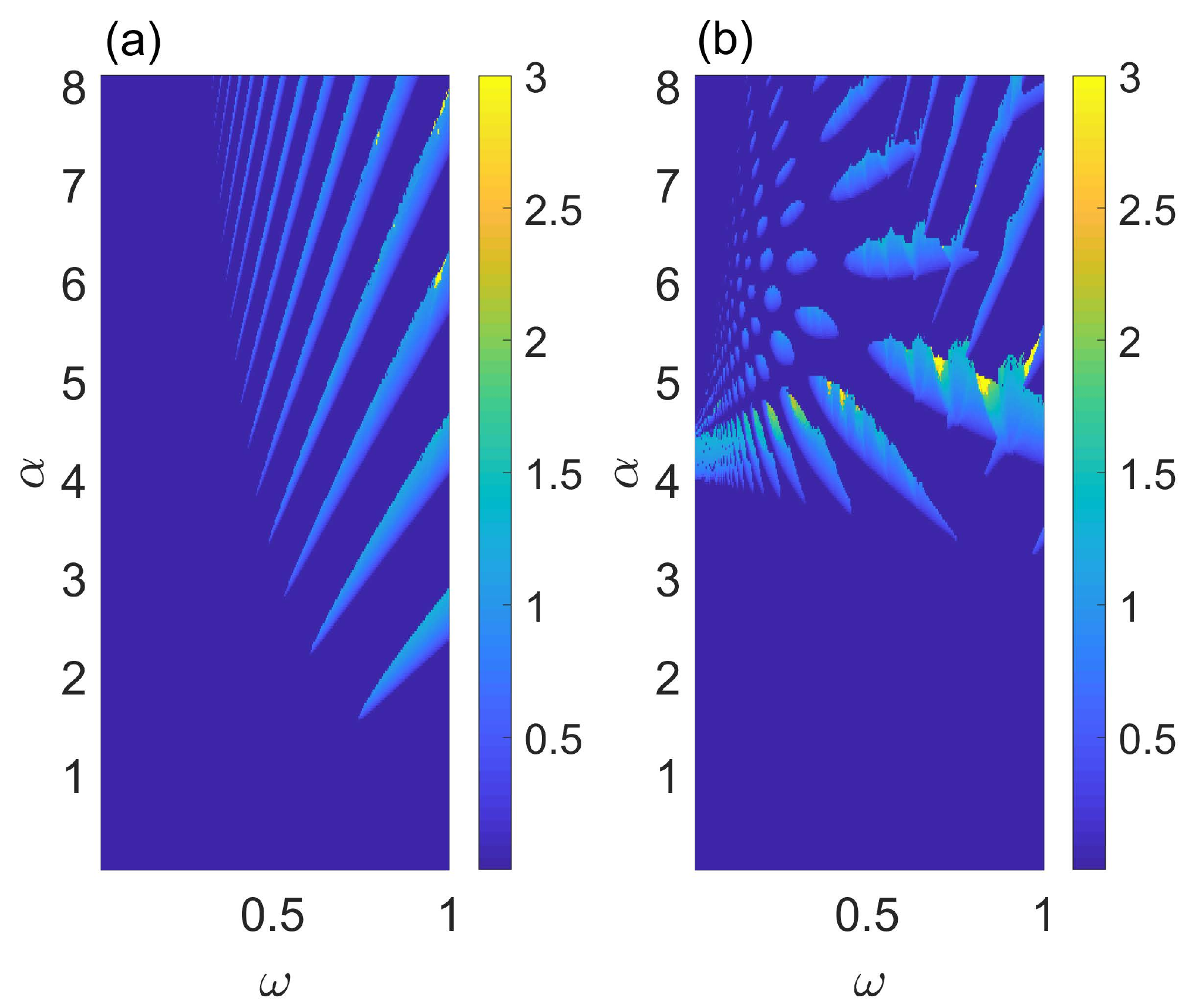}
\caption{\label{dxdfxi}
Deviation of motion $\Delta x$ as a function of strength of sinusoidal ac voltage $\alpha$ and frequency $\omega$ with kicks period $T_k=2\pi/9$. The given parameters are dissipate coefficient $\gamma=0.1$, kicks duration $\varepsilon=10^{-6}$. (a) weak kicks show Arnold tongues which indicate symmetry breaking $\xi=300$. (b) strong kicks develop new instabilities in dynamics $\xi=1300$. }   
\end{figure}

\subsection*{Driven nano-shuttle with kicks}

Now let us turn on the kicks through a discrete pulsed bias voltage. 
The discrete pulses induce a non-smooth function and break the time-translational symmetry of the instantaneous current while the system and applied bias voltage keep the geometrical and temporal symmetries.

Figure \ref{dxdfxi} shows the deviation of the position of the oscillating shuttle $\Delta x$ as a function of strength and frequency of the sinusoidal ac voltage with given kick period and strength, where $\Delta x=x_{max}-x_{min}$, $x_{max}$ is the maximum shuttle position, and $x_{min}$ is the minimum shuttle position.
The quantity $\Delta x$ could tell us the shuttle trajectory decay to balanced zero position or have finite motion region. 
The motion of shuttle shows Arnold tongues which indicate unstable limit circles when the kick strength is weak as shown in fig. \ref{dxdfxi} (a). Each tongue indicates the various oscillation corresponding frequency. The kicks generate self-oscillations in terms of perturbing the stable fixed points. The Features reflect the symmetry breaking of motion as the same of results in Ref. \cite{ahn2006current}. 
When the kick strength is strong, there are more unstable features as shown in fig. \ref{dxdfxi} (b). The strong kicks generates new tongues and lift up original tongues. The main features indicate the time-translational-symmetry breaking and the chaotic behavior.

\begin{figure}
\centering
\includegraphics[width=1.0\linewidth]{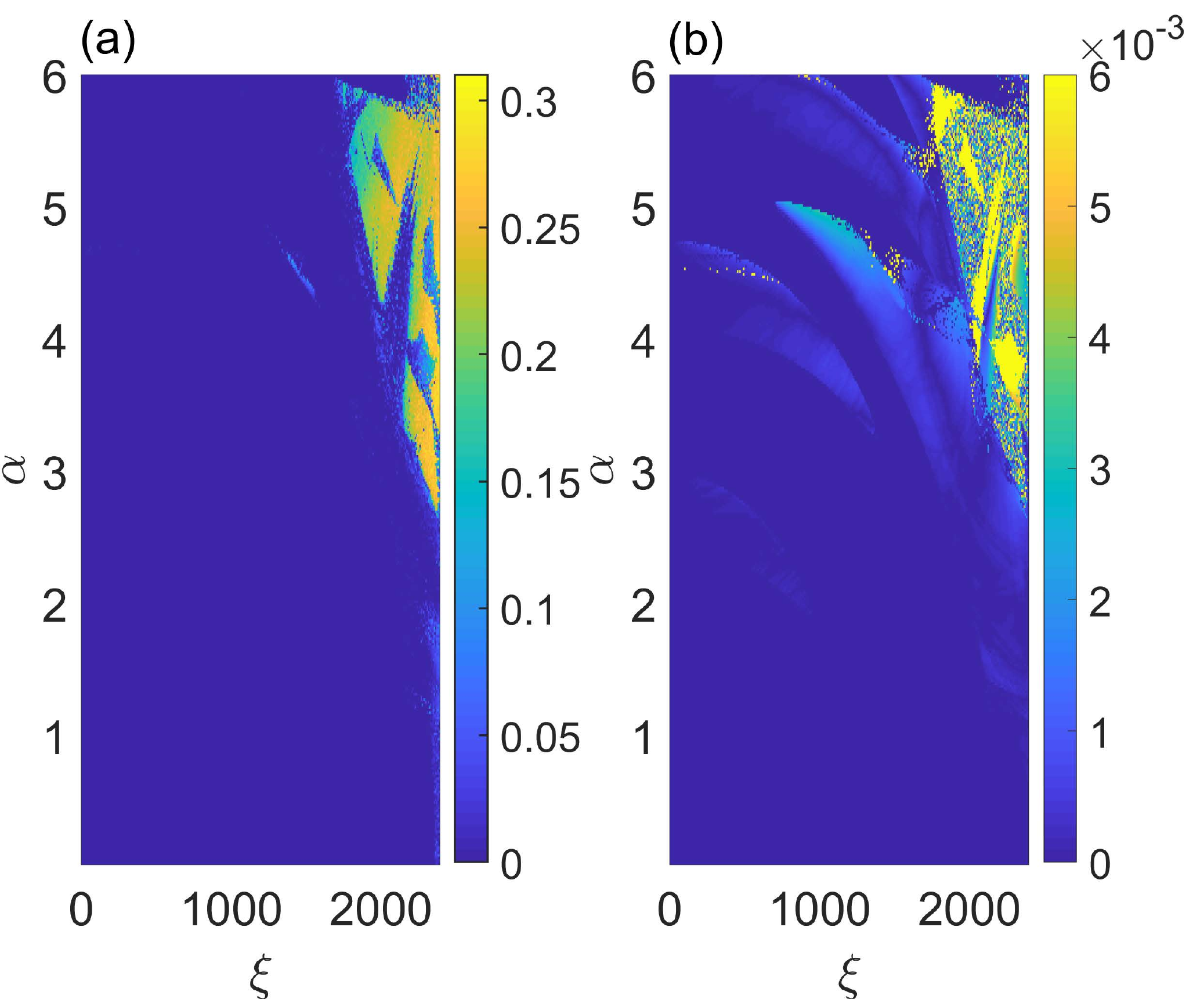}
\caption{\label{lycur}
(a) Lyapunov exponent $\lambda$ and (b) rectified current $I_{dc}$ as a function of strength of sinusoidal ac voltage $\alpha$ and kicks $\xi$. The parameters are the same of parameters in fig. \ref{dxdfxi} with given frequency of sinusoidal ac voltage $\omega=1$. There are finite current in both the regular and the chaotic motion.} 
\end{figure}

In order to elucidate the chaotic motion due to kicks, we introduce the Lyapunov exponent as an important quantity to characterize a chaotic or regular motion of a dynamic system. 
The maximal Lyapunov exponent can be defined as follows,
\begin{equation}
\lambda=\lim_{N\rightarrow\infty}\lim_{\delta Z(0)\rightarrow0}\frac{1}{N}\ln\frac{|\delta Z(N)|}{|\delta Z(0)|},
\end{equation}
where $\delta Z(0)$ is the initial separation of two trajectories in phase space, and $\delta Z(N)$ is the separation of two corresponding trajectories after $N$ kicks period.
Given parameters, the Lyapunov exponent indicates that the motion of shuttle is developed from fixed points or regular motions to chaotic motions as increasing kicks strength and coupling constant in fig. \ref{lycur} (a). 
Fig. \ref{lycur} (b) shows finite the scattered rectified currents in chaos region. 
In addition to the scattered rectified currents by chaotic motions, there are several current tongues caused by regular motion in fig. \ref{lycur} (b) which are absent in fig. \ref{lycur} (a).  
We point out through two figures that the shuttle exhibits fixed points, regular motions, and chaotic motions.
The shuttle system generates corresponding nonzero rectified current in both regular and chaotic regions due to the symmetry breaking.

\subsection*{Role of kicks in driven nano-shuttle}

\begin{figure}
\centering
\includegraphics[width=1.0\linewidth]{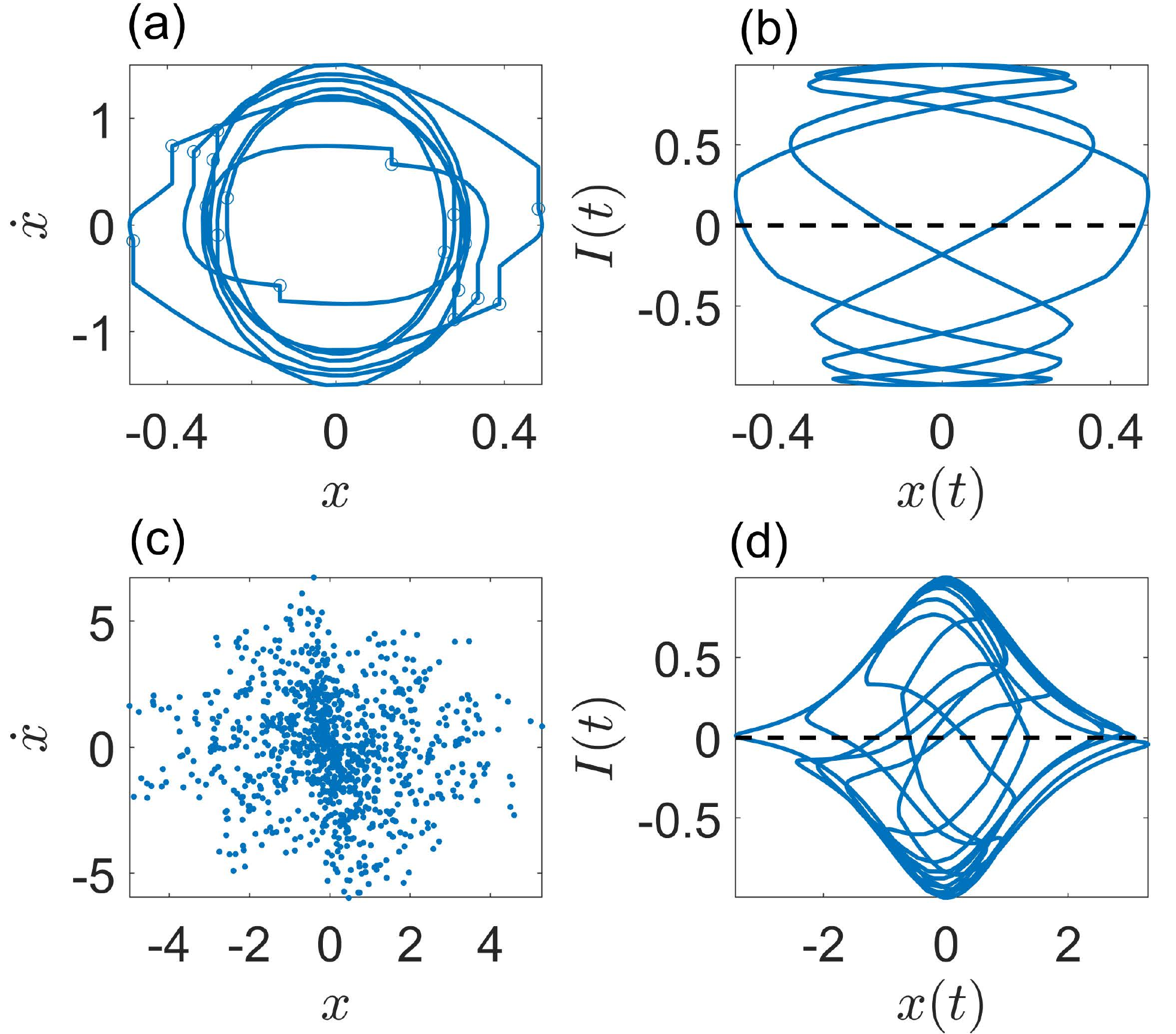}
\caption{\label{trajcurx}
(a), (c) the trajectories in the phase space $(x(t),\dot{x}(t))$ and (b), (d) corresponding instantaneous current as a function of position $(x(t),I(t))$, and (a), (b) are picked at $(\xi,\alpha)=(1050,4.9)$ and (c), (d) are picked at $(\xi,\alpha)=(2000,5.08)$.The parameters are the same of parameters in fig. \ref{lycur}. (a) shows regular motion and (b) gives us instantaneous current of which the time-translational symmetry is broken. (c) shows chaotic motion and (d) gives us aperiodic instantaneous current.}  
\end{figure}

The trajectories in the phase space and corresponding instantaneous currents as a function of position contain fruitful information for finite rectified currents on Arnold tongues, $(x(t),\dot{x}(t))$ and $(x(t),I(t))$, respectively. 
We can select the special points corresponding to the regular motion and chaotic motion in fig. \ref{lycur}. 
In fig. \ref{trajcurx} (a) and (b), the stroboscopic trajectory in phase space shows regular motion and the instantaneous current exhibits time-translational symmetry breaking in one period so that the rectified current is finite. 
In fig. \ref{trajcurx} (c) and (d), meanwhile, the stroboscopic trajectory in the phase space shows scattered as the chaotic motion, and the instantaneous current exhibits irregular pattern because of such chaotic motion.
We can see that the kicks greatly generate the rectified current in the geometrically symmetric electron shuttle in both regular and chaotic region.

The parts of corresponding time-dependent forces $F_c\sim\sin^2\omega t$ and $F_k\sim D_{\varepsilon}(t)$ have the $\pi$ and $2\pi/9$ periods for the calculation, respectively. 
The period for the equation of motion Eq.(\ref{mequ}) is determined the lowest common multiple of periods of the two forces and spatial symmetry. 
The position of the shuttle has $2\pi$ period with $x(t+2\pi)=x(t)$ or $4\pi$ period with $x(t+2\pi)=-x(t)$. The instantaneous current is an odd function of time and an even function of the position as shown in Fig.  \ref{trajcurx} (b). Therefore, the rectified current is finite for the regular motions. 
For the rectified current, the term $\sin\omega t$ keeps the same value under the time shift $t\rightarrow t+2\pi$ (Eq.(\ref{dcc})), there is no time-translational symmetry cancellation during the integration within one period of instantaneous current. The rectified current is determined by the properties of the motion within time interval $[0,2\pi]$. The shuttle now doesn't have parity symmetry $(x\rightarrow -x)$ with time shift $t\rightarrow t+\pi$, therefore the time-translational symmetry of the instantaneous current is broken and the rectified current take nonzero value.

On the other hand, the chaotic current is generated by chaotic motion of the nano-shuttle caused by the periodic kicks. The kicked nano-shuttle can be an analogue of kicked rotors of which strong kicks induce the chaotic phase. The nano-shuttle quasi-randomly transfers electrons through stochastic motions due to the strong kicks.
For two variable expanding method with a small constant $A$, we define two time variables, $\kappa=t$ and $\eta=At$. The equation of motion of nano-shuttle is reduced as (prime is omitted and the equation is dimensionless through re-scaling $t^{\prime}=2\omega t$), 
\begin{equation}
\ddot{x}+\bar{\gamma}\dot{x}+\delta^{2}
x+\mathcal{F}(x)\left[A-A\cos(t)+B D_{\varepsilon}(t)\right]=0
\label{xchaos}
\end{equation}
where $\bar{\gamma}=\gamma/2\omega$, $\delta=1/2\omega$,
$A=\alpha^2/8\omega^{2}$, $B=\xi^{2}/4\omega^{2}$ \cite{kevorkian1966zeipel}.
The time derivative of $x$ can be expressed as the derivative of $\kappa$ and $\eta$.
With expanding $x=x_0+Ax_1+\cdots$, $\delta=\delta_0+A\delta_1+\cdots$, and $\mathcal{F}(x)=\mathcal{F}(x_{0})+A\mathcal{F}^{\prime}(x_{0})x_{1}+\cdots$, we can get following equations after collecting the terms with the same order,
\begin{eqnarray}
&&\frac{\partial^2x_0}{\partial \kappa^{2}} + \delta_{0}^{2}x_0 +
\mathcal{F}(x_0)B D_{\varepsilon}(\kappa) =  0 \label{x0e}\\
&&\frac{\partial^2x_1}{\partial \kappa^{2}} + \delta_{0}^{2}x_1 +
\mathcal{F}^{\prime}(x_{0})x_1 B D_{\varepsilon}(\kappa) \label{x1e}\\
&&=-\mathcal{F}(x_0)(1-\cos\kappa) - 2\frac{\partial^2x_0}{\partial
\kappa\partial\eta} -\mu\frac{\partial x_0}{\partial\kappa}-2\delta_{0}\delta_{1}x_{0},\nonumber
\end{eqnarray}
where $\mu=\bar{\gamma}/A$. 
Let us  expand $\mathcal{F}(x_0) \sim x_{0}$ with small $x_0$. 
Equation (\ref{x0e}) and (\ref{x1e}) are rewritten as follow,
\begin{eqnarray}
&&\frac{\partial^2x_0}{\partial \kappa^{2}} + \delta_{0}^{2}x_0 + x_0 B
D_{\varepsilon}(\kappa)=  0 \label{lx0e}\\
&&\frac{\partial^2x_1}{\partial \kappa^{2}} + \delta_{0}^{2}x_1 + x_1 B D_{\varepsilon}(\kappa) \label{lx1e}\\
&&=-x_0(1-\cos\kappa) - 2\frac{\partial^2x_0}{\partial \kappa\partial \eta} -\mu\frac{\partial x_0}{\partial\kappa} -2\delta_{0}\delta_{1}x_{0}\nonumber
\end{eqnarray}

\begin{figure}
\centering
\includegraphics[width=1.0\linewidth]{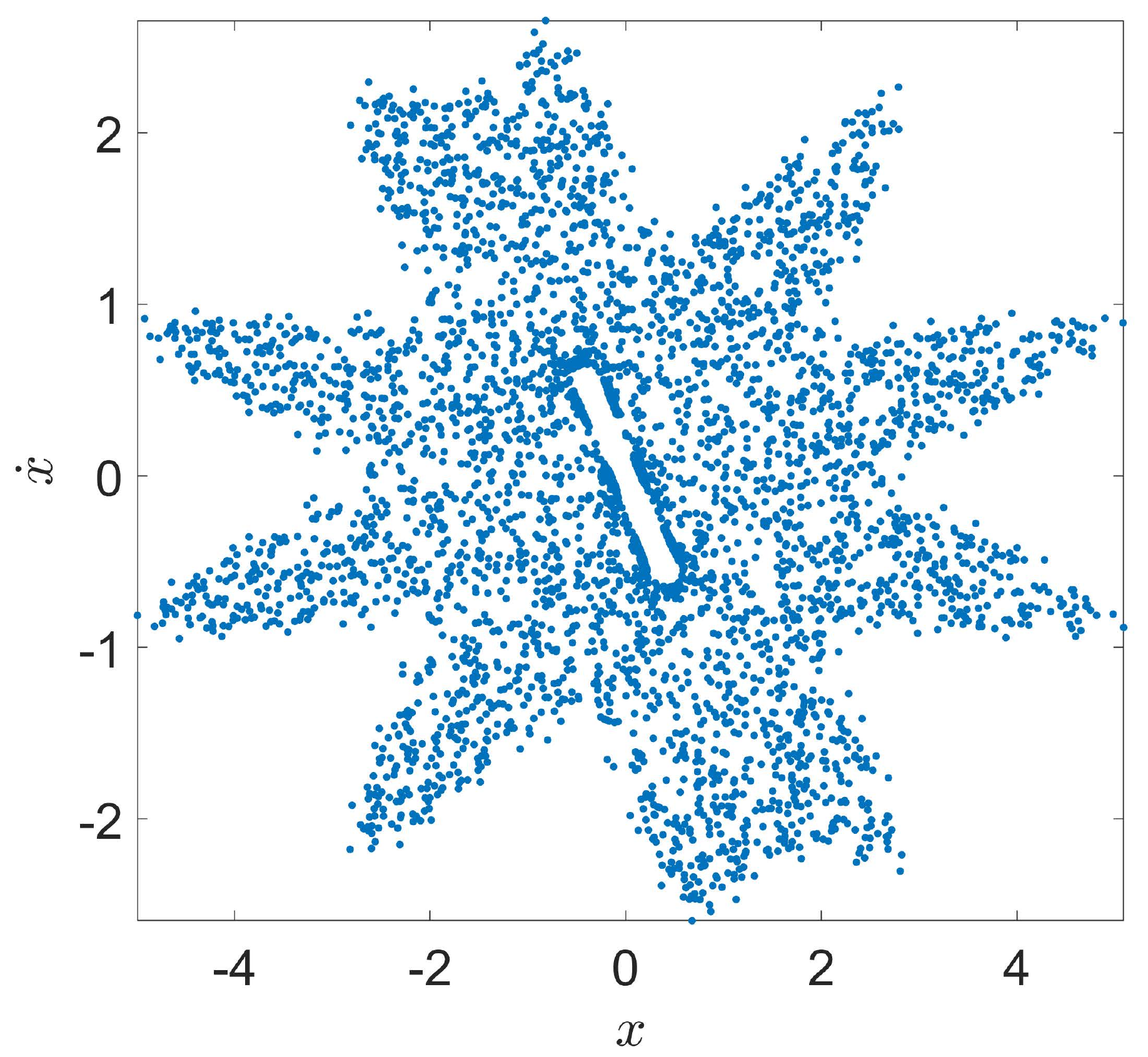}
\caption{\label{fig:org72bedf5}
Trajectory of the shuttle in the phase space $(x(t),\dot{x}(t))$. The figure is calculated by Eq.(\ref{x0e}) whit the parameters ($T_k=4\pi/9$, $\delta_0=0.49$, $B\varepsilon=3$).} 
\end{figure}

We can see that the Eq. (\ref{lx0e}) indicates kicked harmonic oscillator.
The condition of trajectory with the positive Lyapunov exponent is derived as follow,
\begin{equation}
\left|\cos\delta_{0}T_k-B\varepsilon\sin (\delta_{0}T_k)/2\delta_0\right|>1,
\end{equation}
where $T_k$ is the period of $D_{\varepsilon}(t)$ in Eq.(\ref{xchaos}).
In this condition, $x_{0}$ quickly diverges after several steps of time evolution, and destroys the validity of the expansion $\mathcal{F}(x_{0})\sim x_{0}$. 
However, the nonlinearity of $\mathcal{F}(x_{0})$ in Eq.(\ref{x0e}) guarantees to confine the trajectory and chaotic motion of $x_0$ can be possible.
We can show such a chaotic motion by using the numerical calculation.
Figure \ref{fig:org72bedf5} shows the chaotic phase diagram through the Lyapunov exponent with parameters $T_k=4\pi/9$, $\delta_0=0.49$, and $B\varepsilon=3$. It is satisfied by the condition of positive Lyapunov exponent which is $1.16$ and consistence with the chaotic trajectory of Fig. \ref{trajcurx} (c).
As the result of such chaotic motion, the time-translational symmetry of the instantaneous current is broken and the rectified current is finite and quasi-randomized, even though the system preserves the geometrical symmetry.

\subsection*{Conclusion}
In conclusion, we study a geometrically symmetric electron shuttle with kicks and sinusoidal ac voltage. 
If the kick is turned off, there is no rectified current due to the time-translational symmetry of the instantaneous current, even though the shuttle has finite motion displacement between two electrodes. 
However, the kick changes the period of shuttle motion and induces the incommensuracy of the period between shuttle motion and driven bias voltage.
The incommensurate period breaks the time-translational symmetry of the instantaneous current so that the rectified current is finite with the regular shuttle motion. 
In addition, the electron shuttle exhibits chaotic motion due to the nonlinear force caused by the kicks. Under the chaotic motion, the time-translational symmetry of the instantaneous current is obviously broken and the rectified current is nonzero even with geometrically symmetric shuttle.
Our results are discrepant with the general expectation that the zero rectified current of the geometrically symmetric electron shuttle. A plenty of properties of potentially important rectified current are related to a special driven bias voltage and novel applications for the nano-electromechanical systems.

\emph{Acknowledgments}
This work was supported by Project Code (IBS-R024-D1) and the Korea Institute for Advanced Study (KIAS) funded by the Korean Government (MSIT).

\appendix

\section{Derivation of chaotic region}
\label{apa}
The equation of motion for kicked harmonic oscillator is following, see Eq. (\ref{lx0e}),
\begin{equation}
\frac{\partial^2x_0}{\partial \kappa^{2}} + \delta_{0}^{2}x_0 + x_0 B
 D_{\varepsilon}(\kappa)=  0. 
\end{equation} 
Suppose we know the position and momentum of the oscillator
$(x_{0}^n,p_{0}^n)$ at the beginning of time interval
$[nT_{k}-\varepsilon,nT_{k})$, where
$x_{0}^n=x_{0}(nT_{k})$, $p_{0}^n=\dot{x}_{0}(nT_{k})$.
Under the small $\varepsilon$, the position is unchanged at the end of this time period, but the momentum changes into $p_0^n-B\varepsilon x_{0}^n$.
This point on the phase space is the initial state for next time interval $[nT_k,(n+1)T_k-\varepsilon)$. 
In this period, the oscillator behaves exactly as a harmonic oscillator. The position and momentum are determined by,
\begin{eqnarray}
x_0^{n+1} & = & x_0^n \cos \delta_{0}T_k + \frac{\left(p_0^n-B\varepsilon x_0^n\right)}{\delta_{0}} \sin \delta_{0}T_k \label{xnp1}\\
p_0^{n+1} & = & \left(p_0^n-B\varepsilon x_0^n \right) \cos \delta_{0}T_k -\delta_{0} x_0^n \sin \delta_{0}T_k. \label{pnp1}
\end{eqnarray} 
We can get a matrix form as following,
\begin{equation}
\left[ \begin{array}{c} x_0^{n+1}\\ p_0^{n+1} \end{array}\right] = M \left[ \begin{array}{c} x_0^{n}\\ p_0^{n} \end{array}\right],
\end{equation}
where
\begin{equation}
M= \left[
\begin{array}{cc}
\cos \delta_{0}T_k - \frac{B\varepsilon}{\delta_{0}} \sin\delta_{0}T_k & \frac{1}{\delta_{0}} \sin \delta_{0}T_k \\ 
-B\varepsilon \cos \delta_{0}T_k-\delta_{0} \sin \delta_{0}T_k & \cos \delta_{0}T_k 
\end{array}
\right].
\end{equation}
It is straightforward to find the condition satisfying $\det (M)=1$.
This means that the multiple of two eigenvalues of matrix $M$ will be $1$.
We can find the eigenvalue as
\begin{equation}
\lambda_{\pm} =  \cos \delta_{0}T_k-\mathcal{B}\sin \delta_{0}T_k \pm \sqrt{\left(\cos \delta_{0}T_k-\mathcal{B}\sin \delta_{0}T_k\right)^{2}-1}, 
\end{equation}
where $\mathcal{B}= B\varepsilon/2\delta_{0}$.

The eigenvalue equation for $M$ can be written by $U$ as follows,
\begin{equation}
MU= U\lambda
\end{equation}
where $\lambda$ is the eigenvalue matrix of $M$.
We will get,
\begin{equation}
U^{-1}\left[
\begin{array}{c}
x_0^{n}\\
p_0^{n}
\end{array}\right] = \lambda^n U^{-1}\left[
\begin{array}{c}
x_0^{0}\\
p_0^{0}
\end{array}\right]
\end{equation}
Suppose matrix $U^{-1}$ has form as,
\begin{equation}
U^{-1}=\left[
\begin{array}{cc}
\mathcal{U}_a & \mathcal{U}_b \\
\mathcal{U}_c & \mathcal{U}_d
\end{array} \right]
\end{equation}
Then we have,
\begin{eqnarray}
\mathcal{U}_a x_0^{n}+\mathcal{U}_b p_0^{n} & = & (\lambda_{+})^{n} \left( \mathcal{U}_a x_0^{0} +\mathcal{U}_b p_0^{0} \right)\\
\mathcal{U}_c x_0^{n}+\mathcal{U}_d p_0^{n} & = & (\lambda_{-})^{n} \left( \mathcal{U}_c x_0^{0} +\mathcal{U}_d p_0^{0} \right)
\end{eqnarray}
where $\lambda_{\pm}$ are the corresponding eigenvalues.
From this equation set, $0$th position at $n$th discrete time is determined as follow,
\begin{eqnarray}
x_0^{n} &=& \frac{1}{\mathcal{U}_a\mathcal{U}_d-\mathcal{U}_b\mathcal{U}_c}
\left[ (\lambda_{+})^{n}\mathcal{U}_d\left( \mathcal{U}_ax_0^{0}+\mathcal{U}_bp_0^{0} \right)\right. \nonumber\\ 
&&- \left.(\lambda_{-})^{n} \mathcal{U}_b \left( \mathcal{U}_cx_0^{0}+\mathcal{U}_dp_0^{0} \right) \right] 
\label{xsl}
\end{eqnarray}
Suppose we have $|\cos \delta_{0}T_k-\mathcal{B}\sin \delta_{0}T_k|>1$ and
$\lambda_{-}<1$, then $(\lambda_{-})^{n}\rightarrow0$. 
We get
\begin{equation}
x_0^{n} = \frac{\mathcal{U}_d (\lambda_{+})^{n}\left( \mathcal{U}_ax_0^{0}+\mathcal{U}_bp_0^{0} \right) }{\mathcal{U}_a\mathcal{U}_d-\mathcal{U}_b\mathcal{U}_c} = \mathcal{C} e^{n\ln\lambda_{ +}} 
\end{equation}
where $\mathcal{C} = \mathcal{U}_d\left( \mathcal{U}_ax_{0}^0+\mathcal{U}_bp_{0}^0 \right)/(\mathcal{U}_a\mathcal{U}_d-\mathcal{U}_b\mathcal{U}_c)$ is the constant
related to the initial condition $x_{0}^0, p_{0}^0$. 
Now we have positive Lyapunov exponent $L_y=\ln\lambda_+$.

\bibliographystyle{apsrev4-1}
\bibliography{SyElSh}

\end{document}